# Bayesian posterior probabilities: revisited

## David A. Morrison

Section for Parasitology (SWEPAR)
Department of Biomedical Sciences and Veterinary Public Health
Swedish University of Agricultural Sciences, 751 89 Uppsala, Sweden
Email: David.Morrison@bvf.slu.se

**Abstract**

Huelsenbeck and Rannala (2004, Systematic Biology 53, 904–913) presented a series of simulations in order to assess the extent to which the bayesian posterior probabilities associated with phylogenetic trees represent the standard frequentist statistical interpretation. They concluded that when the analysis model matches the generating model then the bayesian posterior probabilities are correct, but that the probabilities are much too large when the model is under-specified and slightly too small when the model is over-specified. Here, I take issue with the first conclusion, and instead contend that their simulation data show that the posterior probabilities are still slightly too large even when the models match. Furthermore, I suggest that the data show that the degree of this over-estimation increases as the sequence length increases, and that it might increase as model complexity increases. I also provide some comments on the authors' conclusions concerning whether bootstrap proportions over- or under-estimate the true probabilities.

# 1 Introduction

Huelsenbeck and Rannala [8] (hereafter HR) presented a series of simulations in order to assess the extent to which the bayesian posterior probabilities associated with phylogenetic trees represent the standard frequentist statistical interpretation, which is that they are the probability of the tree being correct given that the analysis model is correct. Their simulations involved generating aligned sequence data under either the JC69 or the GTR+gamma substitution model and then analysing the data under a variety of substitution models, so that the analysis model either matched the data-generating model, was over-specified (the analysis model being more complex than the generating model) or was under-specified (the analysis model being simpler than the generating model). The authors concluded that when the analysis model matches the generating model then the bayesian posterior probabilities are correct, but that the probabilities are much too large when the model is under-specified and slightly too small when the model is over-specified.

Here, I take issue with HR's conclusion that the bayesian probabilities are correct when the generating and analysis models match, and instead contend that their simulation data show that the posterior probabilities are still slightly too large under these circumstances. Furthermore, I suggest that the data show that the degree of this over-estimation increases as the sequence length increases, and that it might increase as model complexity increases. I also provide some comments on the authors' conclusions concerning whether bootstrap proportions over- or under-estimate the true probabilities.

## 2 Data and Methods

HR presented their results by plotting, for each combination of generating model and analysis model, graphs of the relationship between the true probability (i.e. known from the simulations) and the posterior probability observed from a bayesian analysis of the data. Under these circumstances, if the two probabilities are equal for each simulated data set then the data will form a series of points along the diagonal of the graph (i.e. y=x). However, this approach has long been recognized to be a poor method for comparing two measurements of the same variable on the same experimental units, in spite of its very common use, because it over-estimates the reproducibility of the two measurements [2]. A far better alternative is to produce a mean-difference graph, which is a scatterplot showing the difference between the paired observations on the ordinate (vertically) and their mean on the abscissa (horizontally). This basically involves a rotation and rescaling of the plots shown by HR, so that if the two probabilities are equal for each simulated data set then the data will form a series of points scattered around a horizontal line at y=0. The rotation makes the plot more easily interpreted, because the differences between the measurements are confined to the vertical dimension (rather than being confounded across both dimensions as they are in the standard plot), and the rescaling helps emphasize differences at the equally important small probabilities (which in the standard plot get swamped by the larger probabilities). Such plots have recently become popular in some areas of biology, such as in the analysis of microarray data where they are known as MA plots or R-I graphs [4]; and they have also previously been used in assessments of clade support, such as I am proposing here [e.g. 3].

## 3 Results and Discussion

### 3.1 Bayesian Analysis

In Fig. 1 I present three examples of MA plots, based on the graphs presented by HR. The points on the graphs represent aggregation of the original simulation results into 20 bins based on the posterior probabilities. In each case, I have used the true probability (i.e. known from the simulations) as the abscissa, instead of the average of the true and observed probabilities, as there is assumed to be no sampling variation associated with the true probabilities other than that caused by averaging the data within each bin (cf. a similar approach by [3]). For the ordinate I have used the observed minus true probabilities, so that if the bayesian probabilities over-estimate the true probabilities then the data points will lie above the y=0 line. Fig. 1a shows an example where the analysis model matches the generating model, while Fig. 1b is an example of under-specification, and Fig. 1c is an example of over-specification.

I contend that *all three* graphs show similar patterns, which is not the interpretation provided by HR: when the true probability is close to 0 or 1 then the posterior probability is approximately correct, while it diverges from the true probability at intermediate probabilities. The degree and direction of divergence clearly varies among the three graphs, with under-specification leading to posterior probabilities that are much too large (with a maximum that is nearly twice the true value; Fig. 1b) and over-specification leading to posterior probabilities that are slightly too small (Fig. 1c), as noted by HR. However, it is equally clear that when the analysis model is correctly specified the posterior probabilities are also slightly too large at intermediate probabilities (Fig. 1a), which flatly contradicts the interpretation of HR (p. 908), who state that under these circumstances "the posterior probability of a tree is equal to the probability that the tree is correct". This over-estimate is not large, but a rough heuristic test suggests that the difference between the true and bayesian probabilities is statistically significant (one-sample t-test, $t = 2.3$, $P = 0.031$) — however, it would presumably be better to base such a test on the original data rather than on the summary as represented by the 20 bins shown in the graph.

This over-estimate of the probabilities is also shown by the graphs in Fig. 2. In all three cases the analysis model matches the generating model (GTR+gamma), but the graphs illustrate the effect of increasing the sequence lengths in the simulations. As above, when the true probability is close to 0 or 1 then the posterior probability is approximately correct, while it diverges at intermediate probabilities (c. 0.3–0.9), with the bayesian analysis producing an over-estimate. However, the degree of over-estimation increases as the sequence length increases, with a maximum difference in probability of 0.05 for the longest sequences examined (Fig. 2c). This is not an inconsiderable over-estimation, and the difference between the over-estimates for sequence lengths of 100 (Fig. 2a) and 1000 (Fig. 2c) is statistically significant (two-sample t-test, $t = 2.3$, $P = 0.027$). This simulation is based on only one substitution model and only three sequence lengths, but there seems to be no reason to expect other models or lengths to produce contrary results, and this suggestion is easily (although tediously) tested. Since 1000 nucleotides is only an average-length protein sequence, we might also expect many real data sets to have even stronger over-estimation than is shown in this simulation study.

A comparison of Fig. 1a and Fig. 2a suggests that substitution models with different degrees of complexity might also lead to different degrees of over-estimation by the bayesian posterior probabilities when the analysis model is correctly specified. I base this suggestion on the fact that the maximum degree of over-estimation for the GTR+gamma model (Fig. 2a) is twice that for the JC69 model (Fig. 1a) when compared using the same sequence lengths. However, this remains essentially an untested hypothesis, as the available data are based on only two models and for relatively short sequences. Clearly, it would be interesting to test this hypothesis using data for much longer sequences.

Thus, the data of HR show, when presented in a manner that correctly emphasizes the differences between the two measurements, that the bayesian posterior probabilities over-estimate the true tree probability even when the substitution model used for the data analysis exactly matches the model used to generate the data. Moreover, the degree of over-estimation increases with increasing sequence lengths, reaching an easily detectable level when the sequences are of the average length found in real data sets. It is also possible that the degree of over-estimation increases with increasing substitution-model complexity, so that the over-estimation will be most apparent for the models commonly used to analyse real data sets. However, it is not obvious from the simulation studies of HR to what extent this over-estimation will be problematic in practice for real data sets. Nevertheless, it seems best to assume that bayesian posterior probabilities are overly optimistic in general, unless it can be demonstrated that the analysis model has been over-specified.

Note that these conclusions apply only to the sizes of the data sets used for these simulations. Potential problems of inconsistency in bayesian analyses have been reported for simulated data sets of similar size [9], although this does not necessarily mean that bayesian analysis is inconsistent in general [10].

## 3.2 Boostrap Analysis

HR also calculated non-parametric bootstrap proportions (known as the percentile or naive bootstrap in the statistical literature) for some of their simulations. They concluded from their results that "the bootstrap is too conservative when its assumptions are satisfied" (p. 911). This statement perpetuates a myth that is frequently repeated in discussions of bootstrapping as applied to phylogenetic analysis. However, as clearly shown by the results of the two simulations that they performed (Fig. 3), bootstrap values under-estimate the true probabilities only when the true probability is large, while they *over-estimate* the probability when the true probability is small. That is, the data points quite clearly cross the y=0 line on the graphs, unlike the bayesian posterior probabilities (which form an asymptote to the line), producing over-estimates below the cross-over point and under-estimates above it. This phenomenon was first noted by Zharkikh and Li [13], and it has been repeatedly observed since then (e.g. see the graphs in [1, 5, 6, 7, 11, 12, 14]). However, it seems to have been over-looked by many practitioners, who are lead to believe that *all* bootstrap proportions are under-estimates.

The most problematical aspect of the non-linearity between the true probability of a clade and the bootstrap proportion is that the cross-over between over- and under-estimation does not occur at a fixed probability. The cross-over has been observed to vary in response to: the shape of the tree and the branch lengths, the number of taxa, the number of characters, the evolutionary model used, and the number of bootstrap resamples. For example, Fig. 3 shows the effect of varying the evolutionary model used to generate the data, where under-specification of the analysis model leads to a general over-estimate of the true probability (cross-over at p=0.8; Fig. 3b), while matching the generating and analysis models leads to a general under-estimation (cross-over at p=0.3; Fig. 3a). Also, Fig. 3c illustrates the effect of varying the number of characters. As observed above for the bayesian probabilities, increasing the number of characters leads to an increase in the degree of over- or under-estimation of the true probability, which in this case lowers the cross-over point.

These are potentially serious issues, which seem to be often ignored by practitioners. We can't just assume that the true support value is larger than the observed bootstrap value. In particular, this means that naive bootstrap values are not directly comparable between trees, even for the same taxa, and thus there can be no agreed level of bootstrap support than can be considered to be "significant". A bootstrap value of 90% for a branch on one tree may actually represent *less* support than a bootstrap value of 85% on another tree, depending on the characteristics of the dataset concerned.

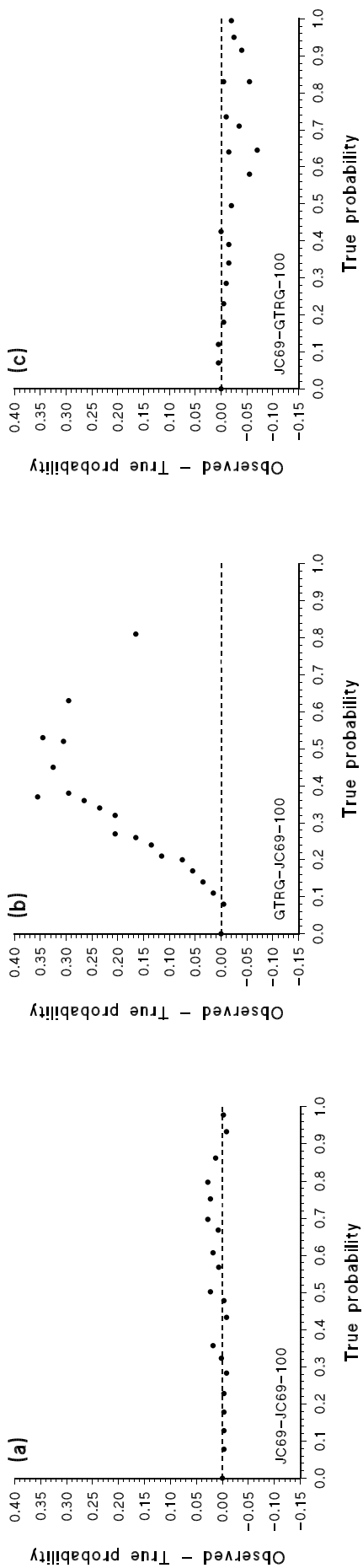

Figure 1. Relationship between the true tree probability and the difference between the observed Bayesian posterior probability and the true probability for three simulated data sets. The label in the bottom corner shows the substitution model used to simulate the data, then the model assumed in the Bayesian analysis, and then the sequence length in nucleotides; JC69 = Jukes-Cantor, GTRG = general time-reversible + gamma-distributed among-site rate variation. The points are based on data presented by [8]. (a) The data-generation model matches the analysis model. (b) The data-generation model is more complex than the analysis model. (c) The data-generation model is simpler than the analysis model.

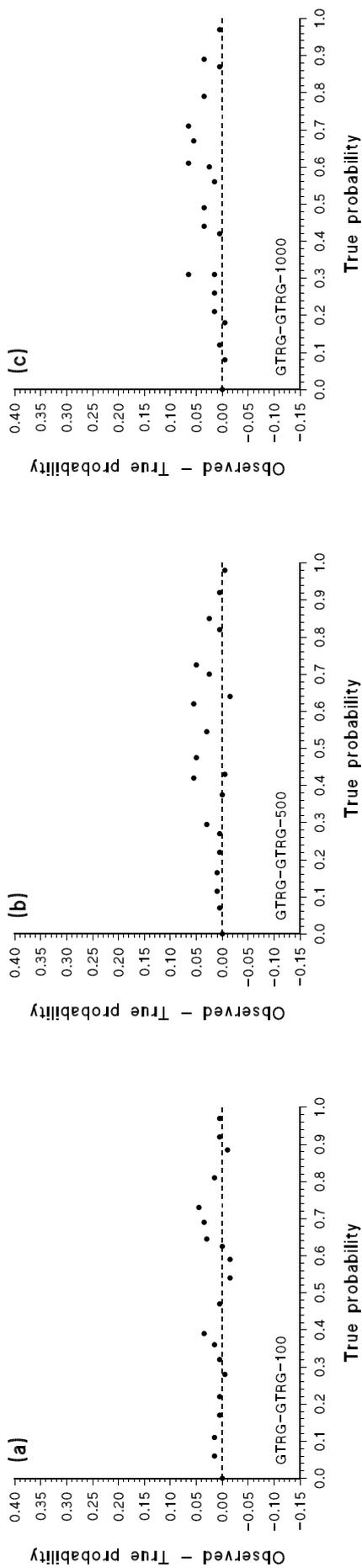

Figure 2. Relationship between the true tree probability and the difference between the observed Bayesian posterior probability and the true probability for three simulated data sets. The label in the bottom corner shows the substitution model used to simulate the data, then the model assumed in the Bayesian analysis, and then the sequence length in nucleotides; GTRG = general time-reversible + gamma-distributed among-site rate variation. The points are based on data presented by [8]. The three simulations differ only in the sequence length generated for the data matrix.

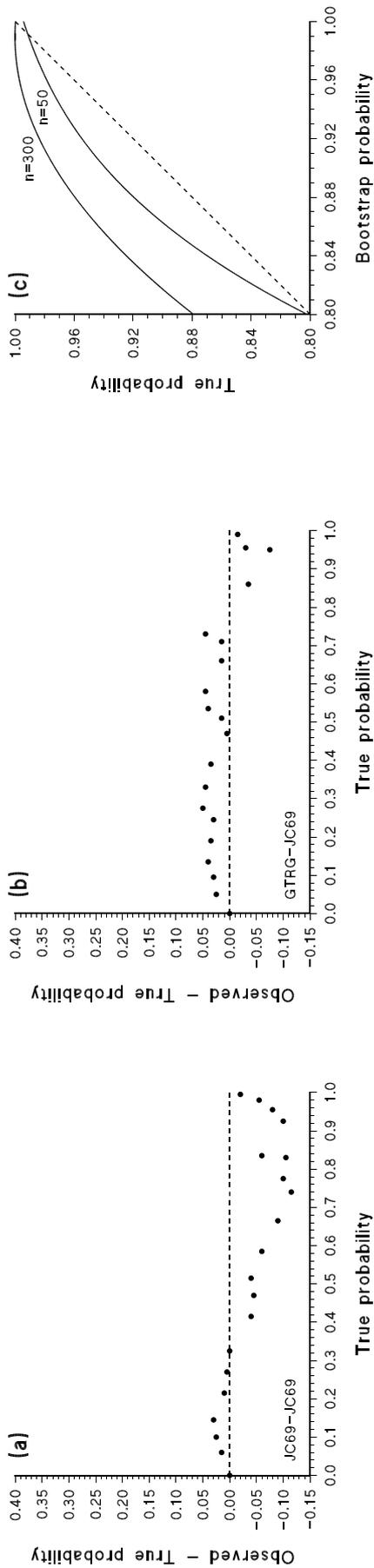

Figure 3. (a) and (b) show the relationship between the true tree probability and the difference between the observed non-parametric bootstrap proportion and the true probability for two simulated data sets. The label in the bottom corner shows the substitution model used to simulate the data, then the model assumed in the bootstrap analysis (the sequence length is 100 nucleotides); JC69 = Jukes-Cantor, GTRG = general time-reversible + gamma-distributed among-site rate variation. The points are based on data presented by [8]. (c) Relationship between the true clade probability and the observed non-parametric bootstrap proportion for two simulated data sets with different numbers of characters (as shown); note the differing axes compared to (a) and (b). The lines are based on data presented by [14] for 1000 bootstrap resamples of a clade of three taxa plus outgroup.